\documentclass[smallextended,epjc3]{svjour3} 
\smartqed  

\RequirePackage{fix-cm}
\RequirePackage{mathptmx}
\RequirePackage{latexsym}
\RequirePackage[numbers,sort&compress]{natbib}
\RequirePackage[colorlinks,citecolor=blue,urlcolor=blue,linkcolor=blue]{hyperref}
\usepackage{color}
\usepackage{dcolumn} 
\usepackage{bm} 
\usepackage{fullpage}
\usepackage{graphicx}
\usepackage{amsmath,amssymb}

\definecolor{dgreen}{cmyk}{1.,0.,1.,0.4} 
\definecolor{orange}{cmyk}{0.,0.353,1.,0.} 

\journalname{Eur. Phys. J. C}

\begin{document}


\title{Event-by-event cumulants of azimuthal angles}

\author{Ante Bilandzic\thanksref{e1,addr1}}

\thankstext{e1}{e-mail: ante.bilandzic@tum.de}

 
\institute{Physik Department, Technische Universit\"{a}t M\"{u}nchen, Munich, Germany \label{addr1}} 
 
\date{Received: date / Accepted: date}

\maketitle

\begin{abstract}
We develop further the recently proposed event-by-event cumulants of azimuthal angles.
The role of reflection symmetry, permutation symmetry, frame independence, and relabeling of particle indices in the cumulant expansion is discussed in detail. 
We argue that mathematical and statistical properties of cumulants are preserved if cumulants of azimuthal angles are defined event-by-event in terms of single-event averages of azimuthal angles, while they are violated in the traditional approach in which cumulants are defined in terms of all-event averages.
We derive for the first time the example analytic solutions for the contribution of combinatorial background in the measured 2- and 3-particle correlations. We demonstrate that these solutions for the combinatorial background are universal as they can be written generically in terms of multiplicity-dependent combinatorial weights and marginal probability density functions of starting multivariate distribution.
The new general results between multiparticle azimuthal correlators and flow amplitudes and symmetry planes are presented.
\keywords{Multivariate correlations and cumulants \and Collective anisotropic flow}
\PACS{25.75.Ld \and 25.75.Gz \and 25.75.-q \and 25.75.Nq}
\end{abstract}

\section{Introduction}
\label{s:introduction}


In ultrarelativistic heavy-ion collisions the properties of an extreme state of nuclear matter, dubbed quark-gluon plasma, can be studied in a controlled environment~\cite{Heinz:2013th,Braun-Munzinger:2015hba,Busza:2018rrf}. Among various experimental observables used to probe its properties, in the present work we focus exclusively on multivariate cumulants~\cite{doi:10.1143/JPSJ.17.1100}. Both at RHIC and LHC experiments they have been regularly applied in the measurements of anisotropic flow phenomenon~\cite{Ollitrault:1992bk,Voloshin:1994mz} and in femtoscopic studies~\cite{Heiselberg:1996xa,Heinz:2004pv}. In the context of flow analyses, the cumulant expansion has been performed traditionally on azimuthal angles~\cite{Borghini:2000sa,Borghini:2001vi,Bilandzic:2010jr,Bilandzic:2013kga}, and only of recently also directly on flow amplitudes~\cite{Mordasini:2019hut,ALICE:2021klf,Bilandzic:2021rgb}. Both the statistical and systematic uncertainties of these measurements are suppressed with inverse powers of multiplicity, which means that multivariate cumulants are a precision tool only in heavy-ion datasets characterized with large values of multiplicity. On the other hand, their measurements in the collisions of smaller objects, like in proton-proton collisions, are unreliable and there is still no consensus in the field on how to interpret them in this environment. 

The mathematical properties of cumulants are rigorous and well established (for the recent concise review and summary, see Ref.~\cite{Bilandzic:2021rgb}). The non-trivial physics information which can be extracted from cumulants stems from the observation that cumulant is identically zero if one of the variables in it is statistically independent of the others~\cite{doi:10.1143/JPSJ.17.1100}. However, this statement is not a mathematical equivalence, and therefore Theorem~1 from Ref.~\cite{doi:10.1143/JPSJ.17.1100} cannot be used to conclude that variables in the cumulant definition are statistically independent of each other if cumulant is zero. Contrary, cumulant is not zero if and only if all variables in it are statistically connected~\cite{doi:10.1143/JPSJ.17.1100}, i.e. there exists a genuine multivariate correlation among all variables in the cumulant, which cannot be written as a superposition of correlations among a smaller number of variables. By definition, each higher-order cumulant extracts a piece of new and independent information that is not accessible at lower orders.

The traditional cumulant expansion over azimuthal angles yields the final expressions that contain only the isotropic all-event averages of azimuthal correlators~\cite{Borghini:2000sa,Borghini:2001vi}. All non-isotropic averages are identically zero due to random fluctuations of the impact parameter vector, for detectors which have uniform azimuthal acceptance~\cite{Borghini:2000sa,Borghini:2001vi}. However, it was recently realized that these results for cumulants of azimuthal angles, in which the non-isotropic azimuthal correlators are missing, do not any longer satisfy the fundamental properties of cumulants~\cite{Mordasini:2019hut,Bilandzic:2021rgb}. The reason for this failure is that mathematical and statistical properties of cumulants are in general preserved only if all terms in the cumulant expansion are kept. As soon as there are underlying symmetries due to which some terms in the cumulant expansion are lost, their mathematical and statistical properties are invalidated, while their physical interpretation can be completely different.

The alternative definition of cumulants of azimuthal angles has been recently proposed in Ref.~\cite{Bilandzic:2021rgb}. Instead of defining cumulants in terms of all-event averages of azimuthal angles, cumulants are defined instead in terms of single-event averages. These new {\it event-by-event cumulants of azimuthal angles} satisfy all mathematical and statistical properties of cumulants. However, their direct applications and measurements are plagued by contributions from the combinatorial background, which are very difficult to remove completely. The standard and approximate mixed-event technique used in high-energy physics to remove combinatorial background is not applicable for azimuthal angles.   

The main motivation for the current draft is an attempt to make further progress with these recently introduced event-by-event cumulants of azimuthal angles. As our main new result, we present how the contributions from the combinatorial background in the measured 2- and 3-particle correlations can be solved analytically, by using the universal combinatorial weights which depend only on multiplicity, and by using marginal probability density functions (p.d.f.). These example solutions can be straightforwardly generalized to correlations involving more than three particles, and for the cases when multiple particles were mis-identified in the calculated correlators---the generalization will be presented in our future work. 

The rest of the paper is organized as follows: In Sec.~\ref{s:role-of-symmetries} we discuss the role of various underlying symmetries which can appear in cumulant expansion. In Sec.~\ref{s:combinatorial-background} we discuss how the presence of combinatorial background affects measurements obtained with correlation techniques.
The new set of general results between multiparticle azimuthal correlators and flow amplitudes and symmetry planes are presented in \ref{a:Expectation-values}.

\section{Role of symmetries}
\label{s:role-of-symmetries}

In this section we discuss the role of a few symmetries which can appear in the cumulant expansion. As our main conclusion, we demonstrate that genuine multivariate correlations can be reliably extracted with cumulants only if there are no underlying symmetries due to which some terms in the cumulant expansion are identically zero. This topic is of great relevance particularly for anisotropic flow analyses in high-energy physics, where most of the cumulants used by now suffer from this problem. In what follows next, we will frequently use the phrase  `p.d.f. does not factorize', with the following meaning: The starting $n$-variate (or joint) p.d.f. $f(x_1,x_2,\ldots,x_n)$ cannot be decomposed into the product of marginal p.d.f.'s corresponding to some set partition of $x_1,x_2,\ldots,x_n$. For instance, factorizable 3-variate p.d.f. would permit the decompositions $f(x_1,x_2,x_3) = f_{{x_1}{x_2}}(x_1,x_2)f_{x_3}(x_3)$, or $f(x_1,x_2,x_3) = f_{x_1}(x_1)f_{x_2}(x_2)f_{x_3}(x_3)$, etc., where $f_{x_{i}x_{j}}$ and $f_{x_i}$ are 2- and single-variate marginal p.d.f.'s of the starting 3-variate p.d.f. $f(x_1,x_2,x_3)$. We use indices for p.d.f.'s to indicate that functional forms of p.d.f.'s can be different, instead of naming them differently. To ease the notation, we drop indices only when we refer to starting $n$-variate p.d.f. of all $x_1,x_2,\ldots,x_n$ variables in question. Theorem~1 from Ref.~\cite{doi:10.1143/JPSJ.17.1100} states that $n$-variate cumulant is identically zero if the underlying $n$-variate p.d.f. is factorizable. Physically, that means that there are no genuine correlations among all $n$ particles in the system, i.e. $n$-particle correlation is just a trivial superposition of correlation involving less than $n$ particles. Mathematically, it is possible to find a set partition of $x_1,x_2,\ldots,x_n$ in which subsets are statistically independent of each other. We now demonstrate that in general the opposite is not true, i.e. cumulant can be identically zero because of underlying symmetries, and not because variables on which the cumulant expansion has been performed are statistically independent. 

\subsection{Reflection symmetry}
\label{s:reflection-symmetry}

We start discussion with the following well-known argument: If $f(x_1,x_2), x_1,x_2\in(-\infty,\infty),$ is a 2-variate p.d.f. which does not factorize, the corresponding two-variate cumulant is not zero. Based on this observation, we conclude that the two variables $x_1$ and $x_2$ are not statistically independent, i.e. there exists a genuine 2-body correlation between them. 
This conclusion fails, however, if the starting p.d.f. $f(x_1,x_2)$ has in addition the following reflection symmetry: $f(x_1,x_2) = f(-x_1,x_2)$. Due to this symmetry, the corresponding 2-variate cumulant, $\langle x_1 x_2\rangle - \langle x_1\rangle\langle x_2\rangle$, is now identically zero, simply because both $\langle x_1\rangle \equiv \int\!\!\int x_1\,f(x_{1},x_{2})\,dx_{1}dx_{2}$ and $\langle x_1x_2\rangle \equiv \int\!\!\int x_1x_2\,f(x_{1},x_{2})\,dx_{1}dx_{2}$ are identically zero. But this result does not imply that $x_1$ and $x_2$ are statistically independent---the starting p.d.f. is still not factorizable. This simple example illustrates clearly the failure of cumulants in the presence of underlying symmetries~\cite{Cowan:1998ji}.

We now generalize this argument to higher-orders. If $f(x_1,x_2,\ldots,x_n), x_1,x_2,\ldots,x_n\in(-\infty,\infty),$ is an $n$-variate p.d.f. which does not factorize, and if
%
%
we again assume that there is a reflection symmetry in one variable $f(x_1,x_2,\ldots) = f(-x_1,x_2,\ldots)$,
it still follows that identically $\langle x_1\rangle = 0$, since:
\begin{eqnarray}
\langle x_1\rangle &\equiv& \int_{-\infty}^{\infty}\cdots\int_{0}^{\infty}x_1f(x_1,x_2,\ldots,x_n)\,dx_1\cdots dx_n\nonumber\\
&=& \int_{-\infty}^{0}\cdots\int_{0}^{\infty}x_1f(x_1,x_2,\ldots,x_n)\,dx_1\cdots dx_n + \int_{0}^{\infty}\cdots\int_{0}^{\infty}x_1f(x_1,x_2,\ldots,x_n)\,dx_1\cdots dx_n\nonumber\\
&=& \int_{\infty}^{0}\cdots\int_{0}^{\infty}(-x_1)f(-x_1,x_2,\ldots,x_n)\,(-dx_1)\cdots dx_n + \int_{0}^{\infty}\cdots\int_{0}^{\infty}x_1f(x_1,x_2,\ldots,x_n)\,dx_1\cdots dx_n\nonumber\\
&=& -\int_{0}^{\infty}\cdots\int_{0}^{\infty}x_1f(-x_1,x_2,\ldots,x_n)\,dx_1\cdots dx_n + \int_{0}^{\infty}\cdots\int_{0}^{\infty}x_1f(x_1,x_2,\ldots,x_n)\,dx_1\cdots dx_n\nonumber\\
&=& -\int_{0}^{\infty}\cdots\int_{0}^{\infty}x_1f(x_1,x_2,\ldots,x_n)\,dx_1\cdots dx_n + \int_{0}^{\infty}\cdots\int_{0}^{\infty}x_1f(x_1,x_2,\ldots,x_n)\,dx_1\cdots dx_n\nonumber\\
&=&0\,.
\end{eqnarray}
The above derivation holds true if we add some of the remaining variables $x_j, x_k, \ldots$ into the correlator, i.e. we have:
\begin{equation}
\langle x_1x_jx_k\cdots\rangle = 0\,,\quad\forall j,k,\ldots\neq 1\,.
\end{equation}
Each term in cumulant expansion is a possible partition of the set $x_1,x_2,\ldots,x_n$, in which in addition the averages have been taken over each subset. Therefore, in each term in the cumulant expansion there is always exactly one average which involves $x_1$, which is identically zero if the underlying $n$-variate p.d.f. has a reflection symmetry $f(x_1,x_2,\ldots) = f(-x_1,x_2,\ldots)$ in that variable. From this we can conclude that due to reflection symmetry in any of the variables, cumulants at any order can be trivial and identical to zero. This does not imply that variables $x_1,x_2,\ldots,x_n$ are statistically independent, instead this is an extreme example in which the cumulant expansion is invalidated at all orders due to underlying symmetries. 

To resolve between the two possibilities, the following additional cross-check needs to be performed in practice: if the underlying $n$-variate p.d.f. factorizes, cumulants are identically zero, irrespective of the choice of sample space for $n$ variables $x_1,x_2,\ldots,x_n$. In this case, cumulants will be identically zero for any choice of boundaries in $x_1\in (x_{1,\rm min},x_{1,\rm max})$, $x_2\in (x_{2,\rm min},x_{2,\rm max})$, etc., and it can be safely concluded that there are no genuine multivariate correlations among all $n$ variables. On the other hand, if cumulants are identically zero due to reflection symmetry, that will occur only for some specific choice of boundaries in $x_1\in (x_{1,\rm min},x_{1,\rm max})$, $x_2\in (x_{2,\rm min},x_{2,\rm max})$, etc.

\subsection{Permutation symmetry}
\label{s:permutation-symmetry}

Next, we consider the role of permutation symmetry in the starting $n$-variate p.d.f. $ f(x_1,x_2,\ldots,x_n)$. For clarity sake, we present the argument for the simplest 2-variate case, since the generalization to higher orders is trivial. 

If for the starting 2-variate p.d.f. we have that $f(x,y) = f(y,x)$, and if the sampling spaces of $x$ and $y$ are identical, it follows immediately that marginal p.d.f.'s are the same, since:
\begin{eqnarray}
f_x(x) &\equiv& \int_y f(x,y)\,dy\nonumber\\
&=& \int_x f(y,x)\,dx\quad {\rm (trivial\ relabelling}\ y\rightarrow x)\nonumber\\
&=& \int_x f(x,y)\,dx\quad {\rm (permutation\ symmetry)}\nonumber\\
&=& f_y(y)\,.
\end{eqnarray}
This conclusion is used frequently later in the paper.

\subsection{Frame independence}
\label{s:frame-independence}

The mandatory requirement which any physics observable needs to fulfill is its independence on the rotations of a coordinate system in the laboratory frame. We can inspect the role of rotations already at the level of 2-particle cumulants of azimuthal angles. In the traditional approach, 2-particle cumulants of azimuthal angles are defined as~\cite{Borghini:2000sa,Borghini:2001vi}:
\begin{eqnarray}
c_n\{2\} &\equiv& \langle\langle e^{in(\varphi_1\!-\!\varphi_2)}\rangle\rangle
-\langle\langle e^{in\varphi_1}\rangle\rangle\langle\langle e^{-in\varphi_2}\rangle\rangle\,.
\label{eq:2p_cumulant_angles_traditional}
\end{eqnarray}
The double angular brackets indicate that in the first step averaging is performed over all distinct combinations of two azimuthal angles  within the same event, and then these results are averaged over all events. If we now rotate randomly by angle $\alpha$ from one event to another the coordinate system in which azimuthal angles are measured, then both $\langle\langle e^{in\varphi_1}\rangle\rangle$ and $\langle\langle e^{-in\varphi_2}\rangle\rangle$ are trivially averaged to zero~\cite{Borghini:2000sa,Borghini:2001vi}. The reason is the following mathematical identity:
\begin{equation}
\langle\langle e^{in(\varphi_1+\alpha)}\rangle\rangle = \langle\langle e^{in\alpha}\rangle\rangle\langle\langle e^{in\varphi_1}\rangle\rangle = 0 \times\langle\langle e^{in\varphi_1}\rangle\rangle = 0\,.
\end{equation}
In practice, the above identity is a direct consequence of random event-by-event fluctuations of the impact parameter vector. Only the first term in Eq.~(\ref{eq:2p_cumulant_angles_traditional}) is invariant with respect to the rotations. 
In this traditional approach, the rotational invariance of cumulants was achieved by simply averaging out all non-isotropic terms. However, by following such a procedure, especially at higher orders, most of the terms in the cumulant expansion are lost, and the final expressions are not any longer the valid cumulants of azimuthal angles.

If we change the definition and instead define cumulants of azimuthal angles event-by-event~\cite{Bilandzic:2021rgb}
\begin{eqnarray}
\kappa_{11} &\equiv& \langle e^{in(\varphi_1\!-\!\varphi_2)}\rangle
-\langle e^{in\varphi_1}\rangle\langle e^{-in\varphi_2}\rangle\,,
\label{eq:2p_cumulant_angles_new}
\end{eqnarray}
the rotational invariance is satisfied by definition, and all terms in the cumulant expansion are kept. We now have that:
\begin{equation}
\langle e^{in(\varphi_1+\alpha)}\rangle\langle e^{-in(\varphi_2+\alpha)}\rangle =
e^{in(\alpha - \alpha)}\langle e^{in\varphi_1}\rangle\langle e^{-in\varphi_2}\rangle = \langle e^{in\varphi_1}\rangle\langle e^{-in\varphi_2}\rangle\,.
\end{equation}
This argument generalizes trivially to higher orders. As our new general result, we argue that  all terms in the cumulant expansion can be kept and simultaneously the rotational invariance of cumulants can be satisfied, only if cumulants of azimuthal angles are defined event-by-event in terms of single-event averages. This conceptual difference in the definition of cumulants of azimuthal angles (all-event averages vs. single-event averages) yields different final results and therefore has non-trivial consequences.

\subsection{Relabeling}
\label{s:relabeling}
Another common practice in the field is to group together in the final expressions for cumulants all azimuthal correlators which are identical, apart from relabeling of indices of azimuthal angles. For instance, in the derivation of the traditional expression for 4-particle cumulant there is the following intermediate result~\cite{Borghini:2000sa,Borghini:2001vi}
\begin{eqnarray}
c_n\{4\} &=& \langle\langle e^ {in(\varphi_1\!+\!\varphi_2\!-\!\varphi_3-\!\varphi_4)}\rangle\rangle\nonumber\\
&-&\langle\langle e^{ in(\varphi_1\!-\!\varphi_3)}\rangle\rangle\langle\langle e^{in(\varphi_2\!-\!\varphi_4)}\rangle\rangle\nonumber\\
&-&\langle\langle e^{in(\varphi_1\!-\!\varphi_4)}\rangle\rangle\langle\langle  e^{in(\varphi_2\!-\!\varphi_3)}\rangle\rangle\,.
\label{eq:4p_cumulant_angles_old_1}
\end{eqnarray}
All 2-particle correlators in the above equation are estimated from the same sample of azimuthal angles. Therefore, they are mathematically identical and to reflect that they are grouped together to obtain the final expression:
\begin{eqnarray}
c_n\{4\} &=& \langle\langle  e^{in(\varphi_1\!+\!\varphi_2\!-\!\varphi_3-\!\varphi_4)}\rangle\rangle
-2\langle\langle e^{in(\varphi_1\!-\!\varphi_2)}\rangle\rangle^2\,.
\label{eq:4p_cumulant_angles_old_2}
\end{eqnarray}
We now argue that such a practice conflicts with the strict mathematical properties of cumulants. In general, $n$-variate cumulant $\kappa_{\nu_1,\ldots,\nu_n}$ is by definition a sum in which each individual summand can be think of as one possible partition of the set consisting of $n$ starting variables $\{x_1,...,x_n\}$~\cite{doi:10.1143/JPSJ.17.1100}. After relabeling variable indices in the final expressions for cumulants, this fundamental property of cumulants is lost. 

This problem is closely related to the general problem of combinatorial background, which we tackle in the next section. Relabeling discussed here amounts essentially to ignoring all the non-trivial effects of combinatorial background. The formalism of cumulants can be applied easily only in cases when combinatorial background plays no role (e.g. in the study of event-by-event fluctuations of different flow magnitudes). However, in cases when the combinatorial background is not under control, the usage of cumulants is not that straightforward.

\section{Combinatorial background}
\label{s:combinatorial-background}

In this section, we investigate the role of combinatorial background in analyses that rely on the usage of correlation techniques and cumulants. We keep the discussion general and therefore our main conclusions are applicable to diverse fields of interest (anisotropic flow, femtoscopy, etc.). 

\subsection{Trivial multiplicity scaling}
\label{ss:Trivial-multiplicity-scaling}

One of the well-known results in high-energy physics is that in the absence of collective phenomena the 2-particle correlators exhibit the trivial multiplicity dependence, the so-called $\sim 1/M$ scaling. In fact, as a very robust signature of collective behavior of produced nuclear matter the breakdown of this scaling is typically proposed. 

We now decipher this trivial multiplicity scaling and demonstrate that it does not originate solely from a few-particle correlations, but rather from the interplay between combinatorial background and a few-particle correlations.  This is a subtle yet important difference. The importance of combinatorial background has been regularly ignored in this context, but we now demonstrate that per se the contributions from a few-particle correlations do not exhibit this trivial multiplicity scaling in the measured 2-particle correlators---the universal nature of $\sim 1/M$ scaling for vastly different physical sources of few-particle correlations stems instead naturally from the presence of combinatorial background. 

In general, in the measurements dominated by a few-particle correlations, the nonflow contribution, $\delta_k$, in the $k$-particle correlator, $\langle\langle k\rangle\rangle$, exhibits the following universal scaling as a function of multiplicity $M$~\cite{Borghini:2000sa,Borghini:2001vi}:
\begin{equation}
\delta_{k} \sim \frac{1}{M^{k-1}}.
\label{eq:nonflowScalingStandard}
\end{equation}
The derivation of this scaling is purely probabilistic and can be derived with the following simple argument. For the sake of simplicity, we outline the argument for 4-particle correlators, but it can be trivially generalized to any higher-order correlator. In an event with multiplicity $M$, we consider four particles which are correlated through direct few-particle correlations which are not of collective origin (for instance, these four particles are originating from the same resonance decay). The probability to select precisely these four particles into the 4-particle correlator, out of $M$ particles available in an event, is given by:
\begin{equation}
\delta_{4} \sim \frac{3}{M-1}\frac{2}{M-2}\frac{1}{M-3} \sim \frac{1}{M^3}.
\label{eq:delta4_scaling}
\end{equation}
After we have taken the first particle into the 4-particle correlator to be the one from the specific 4-particle resonance decay, we have 3 particles left from the same resonance decay out of $M-1$ in total; after we have selected the next one, there are 2 particles left from the same resonance decay out of $M-2$ in total, etc. If we would have instead considered the 5-particle resonance decays but still use 4-particle correlators in the measurements, only the numerical factors in the numerator of Eq.~(\ref{eq:delta4_scaling}) would change, but the multiplicity dependence in the denominator remains the same. As a consequence, the universal multiplicity scaling in Eq.~(\ref{eq:delta4_scaling}) is always present if the measured 4-particle correlators are dominated by contributions from a few-particle correlations, and does not depend on the details of a few-particle correlations.

Only if there are collective effects which induce correlations among all produced particles, the universal multiplicity scaling in Eq.~(\ref{eq:nonflowScalingStandard}) is broken. By following the same probabilistic argument, we have instead that contribution of collective correlations among all $M$ particle in an event, in the mesured $k$-particle correlator is:
\begin{equation}
\delta_{k} \sim 1.
\label{eq:nonflowScalingStandardFlow}
\end{equation}

We remark that their is no room for ambiguity here. For instance, in the context of anisotropic flow analyses, theoretically we can have the fine-tuned possibility that a flow harmonic itself has the following multiplicity dependence:
\begin{equation}
v(M) \sim 1/\sqrt{M}\,.
\label{eq:flowScalingWithM}
\end{equation}
If we now sample all azimuthal angles individually from the single-variate Fourier-like p.d.f. parameterized solely with that harmonic (we use the standard techniques and notation from Ref.~\cite{Bilandzic:2013kga}, see also Sec.~\ref{sss:ToyMCforMscaling}), the 2-particle azimuthal correlator yields:
\begin{equation}
\langle\langle 2 \rangle\rangle = v^2 \sim \frac{1}{M}\,,
\end{equation}
which is the same result as a trivial multiplicity scaling in 2-particle correlators in Eq.~(\ref{eq:nonflowScalingStandard}). The relation between any azimuthal correlator and flow harmonics can be obtained from the general procedure outlined in~\ref{a:Expectation-values}.

To distinguish these two vastly different possibilities, we have to look at higher orders. For 4-particle azimuthal correlator, $\langle\langle 4 \rangle\rangle$, a contribution from a few-particle correlations gives the following generic scaling $\langle\langle 4 \rangle\rangle = \delta_4 \sim 1/M^3$ (see~Eq.~(\ref{eq:nonflowScalingStandard})), while the non-trivial dependence of flow harmonic on multiplicity in Eq.~(\ref{eq:flowScalingWithM}) gives $\langle\langle 4 \rangle\rangle~\sim~1/M^2$. Therefore, we can establish the following argument: if we observe $\langle\langle 2 \rangle\rangle \sim 1/M$, solely based on that measurement we cannot conclude whether the 2-particle correlator is dominated by a few-particle correlations, or by correlations which are of collective origin. However, at higher-orders ($k>2$, $k$ is even) we have the splitting between two possibilities:  
\begin{enumerate}
	\item $\langle\langle k \rangle\rangle \sim 1/M^{k-1}$ $\Rightarrow$ the $k$-particle correlator is dominated by an interplay between a few-particle correlations and combinatorial background;
	\item $\langle\langle k \rangle\rangle \sim 1/M^{k/2}$ $\Rightarrow$ the $k$-particle correlator is dominated by collective effects.
	\label{eq:newFlowSignature}
\end{enumerate}
We have restricted the above argument to even $k$, because for odd $k$ there is an additional contribution from symmetry planes (see Appendix~A), which is not relevant for the discussion presented here. We support this conclusion with the following toy Monte Carlo study.

\subsubsection{The $1/M$ scaling in 2-p correlators: flow or nonflow?}
\label{sss:ToyMCforMscaling}

\begin{figure}
	\begin{center}
		\includegraphics[scale=.5]{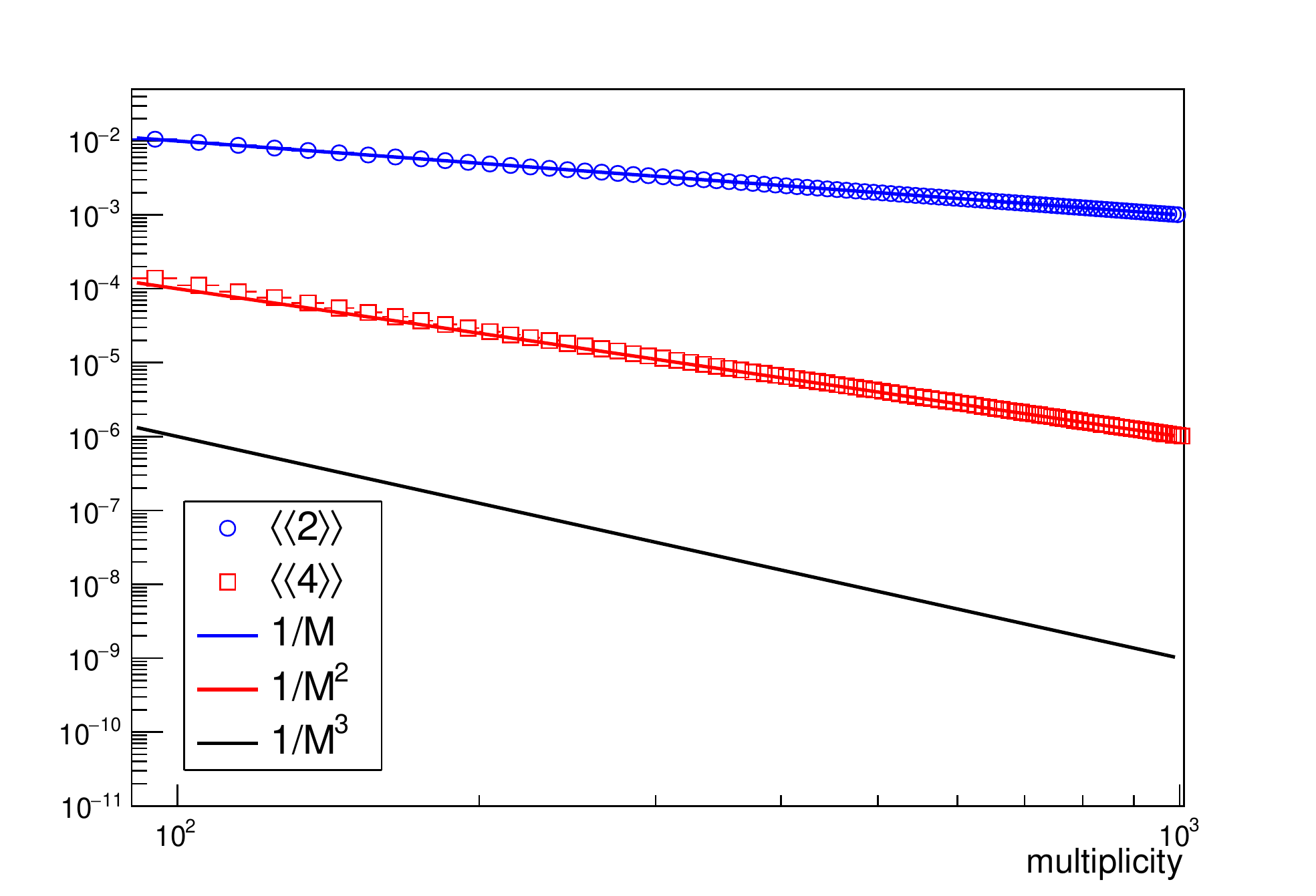}
		\caption{In this study, the elliptic flow $v_2$ itself exhibits $\sim 1/\sqrt{M}$ multiplicity scaling, which yields to $\sim 1/M$ scaling in the 2-particle correlators (blue markers), which is typically attributed to a few-particle correlations. Solely based on the measurement of 2-particle correlator $\langle\langle2\rangle\rangle$ we cannot conclude whether this is a contribution from collective or non-collective effects. However, the 4-particle correlator $\langle\langle4\rangle\rangle$ (red markers) can discriminate between these two possibilities, because it follows what would be the collective flow scaling $1/M^{2}$ in this example (red line), and not a universal few-particle nonflow scaling $1/M^{3}$ (black line). To increase visibility, both horizontal and vertical axes are plotted on log scale.} 
		\label{fig:scaling}
	\end{center}
\end{figure}

As the starting point in this toy Monte Carlo study, we sample the multiplicity $M$ of an event. From the sampled multiplicity, we determine the elliptic flow in that event via $v_2(M) = 1/\sqrt{M}$, so that by definition in this study $\langle\langle 2\rangle\rangle~=~v_2^2~\sim~1/M$, and $\langle\langle 4\rangle\rangle~=~v_2^4~\sim~1/M^2$. In the next step, we sample particle azimuthal angles from Fourier-like single-particle p.d.f. $f(\varphi) = \frac{1}{2\pi}(1 + 2v_2(M)\cos(2(\varphi-\Psi)))$, where $\Psi$ is an event-by-event randomized symmetry plane (this is the standard procedure in the field and it resembles the random event-by-event fluctuations of impact parameter vector). Finally, from the sampled azimuthal angles we calculate the correlators $\langle\langle 2\rangle\rangle$ and $\langle\langle 4\rangle\rangle$ with the Generic Framework~\cite{Bilandzic:2013kga} as a function of multiplicity. The results of this study are presented on Fig.~\ref{fig:scaling}.

\subsection{Example solution for combinatorial background for 2-particle correlations}
\label{ss:General-solution-for-2-particle case}

While the discussion in the previous section on multiplicity scaling in the measured correlators was only qualitative and purely phenomenological, we now provide the robust mathematical treatment. In particular, we establish the mathematical procedure to obtain the analytic solution for the combinatorial background for the example process in which two particles are emitted pair-wise from the most general joint 2-variate p.d.f. $f_{xy}(x,y)$. In the next section we present the solution for 3-particle correlations, in order to demonstrate that the same mathematical procedure is applicable at higher orders. For clarity sake, we consider only the most important cases of particle mis-identification in the measured correlators---the general solutions which cover all possible cases of particle mis-identification will be presented in our future work.

The starting problem can be formulated mathematically as follows:
\begin{enumerate}
	\item From the general 2-variate p.d.f. $f_{xy}(x,y)$ we sample particles in pairs. We perform $M/2$ samplings to obtain an event with multiplicity $M$;
	\item We randomize the final sample of $M$ particles;
	\item What is the p.d.f. $w(x,y)$ of randomized sample, if the starting p.d.f. is $f_{xy}(x,y)$?
\end{enumerate}
We solve the problem for the most general case of starting p.d.f. $f_{xy}(x,y)$, i.e. we do not allow simplifications that can occur due to its factorization $f_{xy}(x,y)= f_{x}(x)f_{y}(y)$, permutation symmetry $f_{xy}(x,y) = f_{xy}(y,x)$, etc. The observables $x$ and $y$ are in general taken from different sample spaces and in general have different functional forms for marginal p.d.f.'s $f_{x}(x)$ and $f_{y}(y)$. To make a connection with experiment, we can tag the starting p.d.f. $f_{xy}(x,y)$ as a `signal', while the p.d.f. $w(x,y)$ after randomization as a `signal + background'. The general 2-particle correlator which corresponds only to signal is $\langle xy\rangle_{S} =\int\!\!\int\! xy\,f_{xy}(x,y)\,dxdy$, while the one that corresponds to both signal and background is $\langle xy\rangle_{S+B} =\int\!\!\int\! xy\,w(x,y)\,dxdy$.

We now demonstrate that the relation between $f_{xy}(x,y)$ and $w(x,y)$ is universal, in a sense that whatever the starting functional form of $f_{xy}(x,y)$ is, the same generic equation determines the functional form of $w(x,y)$. That equation involves only combinatorial weights that depend only on multiplicity $M$, and marginal p.d.f.'s of $f_{xy}(x,y)$. For instance, the marginal p.d.f. $f_{x}(x)$ is the probability to obtain $x$ whatever the value of $y$, and it can be obtained from the following definition $f_{x}(x)\equiv \int\!f_{xy}(x,y)\,dy$. Further discussion about fundamental properties of marginal p.d.f.'s can be found in Ref.~\cite{Cowan:1998ji}. 

\begin{figure}
	\begin{center}
		\includegraphics[scale=.5]{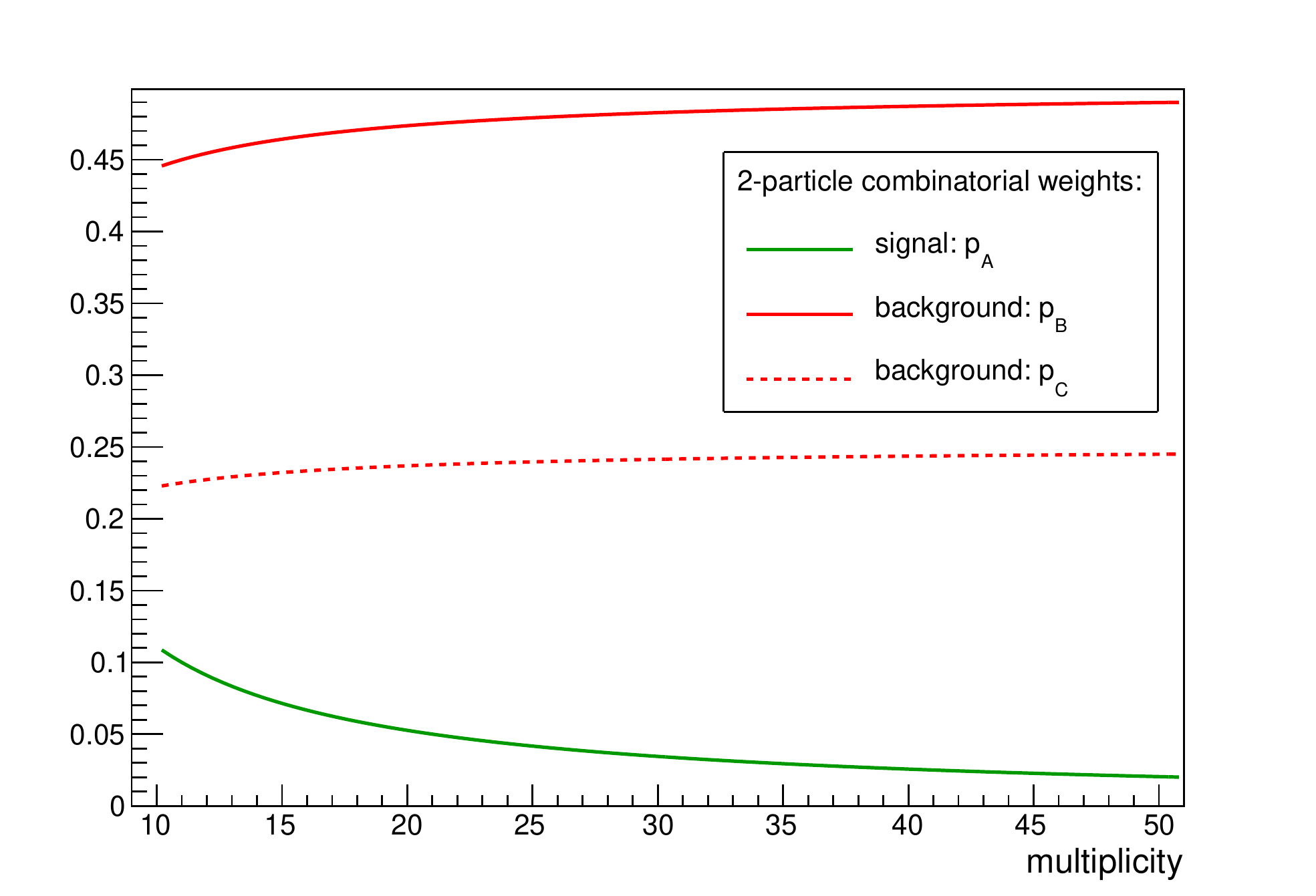}
		\caption{Multiplicity dependence of 3 distinct combinatorial weights, $p_A$, $p_{B}$ and $p_{C}$ (see the main text for their definitions and explanation), 
			which always appear in 2-particle correlations. The contribution from genuine 2-particle correlations (signal) is weighted with $p_A$ (green line). For large $M$ this contribution is suppressed when compared to two distinct contributions from combinatorial background (red lines). The results from this figure cover any case in which particles are produced in pairs from the general 2-variate p.d.f. $f_{xy}(x,y)$.} 
		\label{fig:combinatorialWeights-2p}
	\end{center}
\end{figure}

When particles are sampled pair-wise, and when only the most important case of particle mis-identification is considered, without loss of generality we can divide the final dataset of $M$ particles after randomization in the following 4 disjoint subsets:
\begin{enumerate}
	\item \textbf{diagonal:} both particles originate from the same sampling. Only these two particles can be physically correlated, i.e. this is the signal and is therefore described with the starting p.d.f. $f_{xy}(x,y)$. Its combinatorial weight is $p_A = \frac{2!\frac{M}{2}}{M(M-1)}$;
	\item \textbf{off-diagonal:} two particles are from different samplings and their types were identified correctly. These two particles are statistically independent and therefore their statistical properties are described by the product of two marginal p.d.f.'s. $f_{x}(x)f_{y}(y)$. The combinatorial weight of this contribution is $p_B = 2\frac{\frac{M^2}{4}-\frac{M}{2}}{M(M-1)}$;
	\item \textbf {mis-identified:} to account for the fact that identical particles are indistinguishable, we must allow the possibility that the particle which was originally sampled as $y$ was reconstructed as $x$, and vice versa. For instance, if two pions are emitted from two different processes, after randomization we do not know even in principle which pion originated from which process. This case is governed by 2 possibilities: $f_{x}(x)f_x(y)$ and $f_{y}(y)f_y(x)$. In both cases the combinatorial weight is the same and it reads $p_C = \frac{\frac{M^2}{4}-\frac{M}{2}}{M(M-1)}$\,.
\end{enumerate}
In the above expressions, $f_x$ and $f_y$ are marginal p.d.f.'s of $x$ and $y$, respectively. In total, there are 4 disjoint subsets and we can immediately as a cross-check see that for their respective combinatorial weights it follows:
\begin{equation}
\sum_i^4 p_i = p_A + p_B + 2p_C =  \frac{2!\frac{M}{2} + 2(\frac{M^2}{4}-\frac{M}{2}) + 2(\frac{M^2}{4}-\frac{M}{2})   }{M(M-1)} = 1\,.
\end{equation}
We can now write the example solution for the p.d.f. of randomized sample:
\begin{eqnarray}
w(x,y) &=& p_A f_{xy}(x,y) + p_B f_{x}(x)f_{y}(y) + p_C\big[ f_{x}(x)f_{x}(y) + f_{y}(x)f_{y}(y)\big]\,.
\label{eq:generalSolution2-p}
\end{eqnarray}
The combinatorial weights $p_A$, $p_B$ and $p_C$ are universal---they depend only on multiplicity $M$ and not on any details of the starting p.d.f. $f_{xy}(x,y)$. The universal multiplicity dependence of these 2-particle combinatorial weights in presented in Fig.~\ref{fig:combinatorialWeights-2p}. We stress that the example analytic result in Eq.~(\ref{eq:generalSolution2-p}) can be further generalized by including the contributions when mis-identified particles corresponding to the same initial particle type are coupled directly to each other in the 2-particle correlators.  

The practical use case of the above result amounts to the fact that we can now decompose and study separately the individual contributions to the measured 2-particle correlators in the randomized sample, and deduce their relative importance. For instance, the standard 2-particle azimuthal correlation used in flow analyses can be calculated in the randomized sample as follows:
\begin{equation}
\langle 2\rangle \equiv \langle\cos[n(\varphi_1\!-\!\varphi_2)]\rangle = \int\!\!\!\!\int\cos[n(\varphi_1\!-\!\varphi_2)]w(\varphi_1,\varphi_2) \,d\varphi_1 d\varphi_2\,,
\end{equation}
where the p.d.f. $w(x,y)$ is given by Eq.~(\ref{eq:generalSolution2-p}), while the azimuthal angles $\varphi_1$ and $\varphi_2$ can in general be taken from different sample spaces. The above equation describes analytically what is measured experimentally when we use 2-particle azimuthal correlations, for the case when particles are emitted pair-wise, when the final dataset is randomized, and when only the most important cases of particle mis-identification are taken into account. Given the nature of the example expression for the p.d.f. $w(x,y)$ in Eq.~(\ref{eq:generalSolution2-p}), we are naturally led to decompose distinct individual contributions to 2-particle correlator as follows:
\begin{equation}
\langle 2\rangle = \langle 2\rangle_{w_1} + \langle 2\rangle_{w_2} + \langle 2\rangle_{w_3} + \langle 2\rangle_{w_4}\,,
\end{equation} 
where 
\begin{eqnarray}
\langle 2\rangle_{w_1} &\equiv& p_A \int\!\!\!\!\int\cos[n(\varphi_1\!-\!\varphi_2)] f_{\varphi_1\varphi_2}(\varphi_1,\varphi_2) \,d\varphi_1 d\varphi_2\,,\nonumber\\
\langle 2\rangle_{w_2} &\equiv& p_B \int\!\!\!\!\int\cos[n(\varphi_1\!-\!\varphi_2)] f_{\varphi_1}(\varphi_1)f_{\varphi_2}(\varphi_2) \,d\varphi_1 d\varphi_2\,,\nonumber\\
\langle 2\rangle_{w_3} &\equiv& p_C \int\!\!\!\!\int\cos[n(\varphi_1\!-\!\varphi_2)] f_{\varphi_1}(\varphi_1)f_{\varphi_1}(\varphi_2) \,d\varphi_1 d\varphi_2\,,\nonumber\\
\langle 2\rangle_{w_4} &\equiv& p_C \int\!\!\!\!\int\cos[n(\varphi_1\!-\!\varphi_2)] f_{\varphi_2}(\varphi_1)f_{\varphi_2}(\varphi_2) \,d\varphi_1 d\varphi_2\,,
\end{eqnarray} 
and $f_{\varphi_1}(\varphi_1)\equiv\int f_{\varphi_1\varphi_2}(\varphi_1,\varphi_2)\,d\varphi_2$ and analogously $f_{\varphi_2}(\varphi_2)\equiv\int f_{\varphi_1\varphi_2}(\varphi_1,\varphi_2)\,d\varphi_1$. We illustrate and support this general decomposition with the toy Monte Carlo study in the next section.

\subsubsection{Toy Monte Carlo study for 2-particle case}
\label{sss:Toy-Monte-Carlo-study-CB-2p}
We test the validity of Eq.~(\ref{eq:generalSolution2-p}) with the following clear-cut toy Monte Carlo study, which can be solved analytically. The starting normalized two-variate p.d.f. is:
\begin{equation}
f(\varphi_1,\varphi_2;M) = \frac{9375 \left(\varphi_1-\frac{M \varphi_2^2}{100}\right)^2}{4 \pi ^4 \left(3 \pi ^2 M^2-250 \pi  M+12500\right)}\,,
\label{eq:toy-model-2-variate}	 
\end{equation}
where $M$ is a parameter and it corresponds to multiplicity. The stochastic variables are azimuthal angles $\varphi_1$ and $\varphi_2$, whose sample space is $[0,2\pi)$. One can easily check that:
\begin{equation}
\int_0^{2\pi}\!\!\!\!\int_0^{2\pi}f(\varphi_1,\varphi_2;M)\,d\varphi_1 d\varphi_2 = 1\,.
\end{equation}
for any value of multiplicity $M$.

\begin{figure}[h!]
	\begin{center}
		\includegraphics[scale=.5]{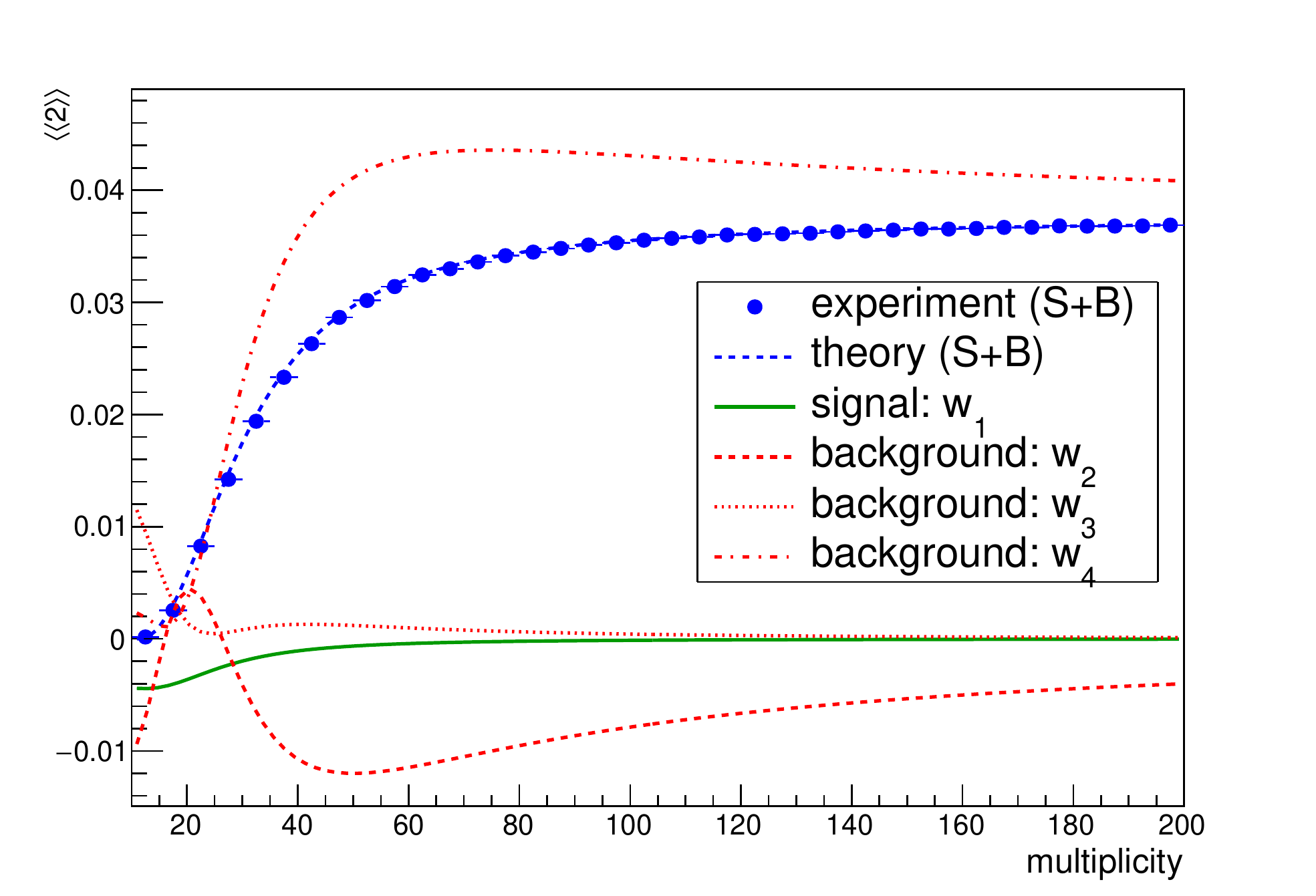}
		\caption{Analytic description of combinatorial background in 2-particle correlations, when particles are produced in pairs, and the final data sample is randomized.}
		\label{fig:toyMC-2p}
	\end{center}
\end{figure}

With straightforward calculus, and after recalling that the combinatorial weight of signal contribution for 2-particle case is $p_A = \frac{2!\frac{M}{2}}{M(M-1)}$, we have obtained analytically for flow harmonic $n=2$ for the signal contribution:
\begin{eqnarray}
\langle 2\rangle_{w_1}
&=&p_A\int_0^{2\pi}\!\!\!\!\int_0^{2\pi}\cos[2(\varphi_1-\varphi_2)] f(\varphi_1,\varphi_2;M) \,d\varphi_1 d\varphi_2\nonumber\\
&=&-\frac{375 M}{4 \pi  (M-1) (\pi  M (3 \pi  M-250)+12500)}\,,
\label{eq:signal-2p}	
\end{eqnarray}
and similarly for the three distinct background contributions $\langle 2\rangle_{w_2}$, $\langle 2\rangle_{w_3}$ and $\langle 2\rangle_{w_4}$. 
%
%
%
%
%
%
All results are shown in Fig.~\ref{fig:toyMC-2p}. With blue markers the experimental result is shown, i.e. the average $\langle 2\rangle = \langle\cos[2(\varphi_1\!-\!\varphi_2)]\rangle$ measured in the randomized data sample, after particle azimuthal angles were sampled pair-wise from p.d.f. in Eq.~(\ref{eq:toy-model-2-variate}). The analytic expression for the signal contribution from Eq.~(\ref{eq:signal-2p}) is shown with the green curve. Red curves represent analytic results for three distinct cases of combinatorial background contribution, $\langle 2\rangle_{w_2}$, $\langle 2\rangle_{w_3}$ and $\langle 2\rangle_{w_4}$, respectively. Finally, the blue curve is a superposition of signal (green curve) and distinct background contributions (three red curves).

\subsection{Example solution for combinatorial background for 3-particle correlations}
\label{ss:General-solution-for-3-particle case}

In this section we address the combinatorial background for 3-particle correlations. The problem we try to solve can be mathematically formulated as follows:
\begin{enumerate}
	\item From the general 3-variate p.d.f. $f_{xyz}(x,y,z)$ particles are emitted in triples. We perform $M/3$ samplings to obtain an event with multiplicity $M$;
	\item We randomize the final sample of $M$ particles;
	\item What is the p.d.f. $w(x,y,z)$ of randomized sample, if the starting p.d.f. is $f_{xyz}(x,y,z)$?
\end{enumerate}
Physically, $f_{xyz}(x,y,z)$ describes `signal', while $w(x,y,z)$ describes `signal + background'. 

For instance, if we are interested in the average 3-particle correlation $\langle xyz\rangle$, then the signal contribution is given by $\langle xyz\rangle_{S}=\int\!\!\int\!\!\int\! xyz\, f_{xyz}(x,y,z)\, dxdydz$, while the contribution from both signal and background is given by $\langle xyz\rangle_{S+B}=\int\!\!\int\!\!\int\! xyz\, w(x,y,z)\, dxdydz$. Given the fact that we have only randomized the initial sample, there should be no systematic biases involved, i.e. we seek the universal relation between $w(x,y,z)$ and $f_{xyz}(x,y,z)$. The derivation of that universal relation is the main result of this section---its generic structure deciphers the role of combinatorial background after we have randomized the final data sample for any starting p.d.f. $f_{xyz}(x,y,z)$. 

We keep the discussion both general and as close to experiment as possible. For instance, we introduce particle mis-identification by blinding the particle labels $x$, $y$, $z$ in the final data sample. This corresponds to the real-case scenario when for instance 3 pions are emitted from three different physical processes. After randomization, even in principle we cannot deduce which pion originated from which process, since identical particles are indistinguishable. Just like in the 2-particle example, we consider here only this most important case of particle mis-identification. The more general solutions which cover all possible cases of particle mis-identification will be presented in our future work. 

If particles are sampled in triples from $f_{xyz}(x,y,z)$, and if we consider only the most important case of particle mis-identifications, without loss of generality we can always divide the sample space after randomization in the following 3 disjoint categories:
\begin{enumerate}
	\item \textbf{diagonal:} all 3 particles are from the same sampling. This case is governed by $f_{xyz}$, i.e. this is the signal. The combinatorial weight of this contribution is $p_{A} = \frac{3!\frac{M}{3}}{M(M-1)(M-2)}$;
	\item \textbf{semi-diagonal:} only 2 particles are from the same sampling. In order to write down p.d.f.'s which govern this case, we have to differentiate further into two disjoint subcategories with different statistical properties. Their corresponding p.d.f.'s and combinatorial weights are:
	\begin{enumerate}
		\item 
		$f_{xy}f_{x}$, $f_{xy}f_{y}$, $f_{xz}f_{x}$, $f_{xz}f_{z}$, $f_{yz}f_{y}$ and $f_{yz}f_{z}$. These cases correspond to the situation when two particles of different type originate from the same sampling. To account for the possibility that there is a genuine 2-particle correlation among them, we use marginal 2-particle p.d.f.'s. The 3rd particle is of the same type, but it originated from separate sampling. In total, there are 6 different possibilities, and the combinatorial weight of each is $p_{B_{1}} = \frac{3!\frac{M}{3}(\frac{M}{3}-1)}{M(M-1)(M-2)}$;
		\item $f_{xy}f_{z}$, $f_{xz}f_{y}$, and $f_{yz}f_{x}$. This is similar to the previous case, the only difference is that the 3rd particle is of different type. There are 3 different possibilities and the combinatorial weight of each is $p_{B_{2}} = \frac{3!\frac{M}{3}(\frac{M}{3}-1)}{M(M-1)(M-2)}$;
	\end{enumerate}
	\item \textbf{off-diagonal:} all 3 particles are from different samplings. To write down the universal decomposition due to combinatorial background in this sector, we have to decompose further into three disjoint subcategories:
	\begin{enumerate}
		\item $f_{x}f_{x}f_{x}$, $f_{y}f_{y}f_{y}$, $f_{z}f_{z}f_{z}$. This is the case when in the 3-particle correlation we have 3 particles of same type, each of which originated from separate sampling. There are 3 distinct cases, and combinatorial weight of each is $p_{C_{1}} = \frac{\frac{M}{3}(\frac{M}{3}-1)(\frac{M}{3}-2)}{M(M-1)(M-2)}$;
		\item $f_{x}f_{x}f_{y}$, $f_{x}f_{x}f_{z}$, $f_{y}f_{y}f_{x}$, $f_{y}f_{y}f_{z}$, $f_{z}f_{z}f_{x}$, $f_{z}f_{z}f_{y}$. In this subcategory, we have 3 particles from 3 different samplings, two of which are from the same type. There are 6 distinct possibilities, and combinatorial weight of each is $p_{C_{2}} = \frac{\binom{3}{2}\frac{M}{3}(\frac{M}{3}-1)(\frac{M}{3}-2)}{M(M-1)(M-2)}$;
		\item $f_{x}f_{y}f_{z}$. Finally, this case corresponds to situation when we have 3 different particles from 3 different samplings in the 3-particle correlator. The combinatorial weight is $p_{C_{3}} = \frac{3!\frac{M}{3}(\frac{M}{3}-1)(\frac{M}{3}-2)}{M(M-1)(M-2)}$.
	\end{enumerate}
\end{enumerate}
In total, in this example there are 20 disjoint cases to deal with, 1 from signal and 19 from combinatorial background.  We can check immediately that the corresponding probabilities sum up correctly:
\begin{eqnarray}
\sum_i^{20} p_i &=& p_{A} + 6p_{B_{1}} + 3p_{B_{2}} + 3p_{C_{1}} + 6p_{C_{2}} + p_{C_{3}}\nonumber\\
&=&\frac{2M + 9\cdot 6\frac{M}{3}(\frac{M}{3}-1) 
	+ (3+6\binom{3}{2}+3!)\frac{M}{3}(\frac{M}{3}-1)(\frac{M}{3}-2)}{M(M-1)(M-2)}\nonumber\\ 
&=& 1\,.
\label{eq:crossCheck}
\end{eqnarray}
\begin{figure}[h!]
	\begin{center}
		\includegraphics[scale=.5]{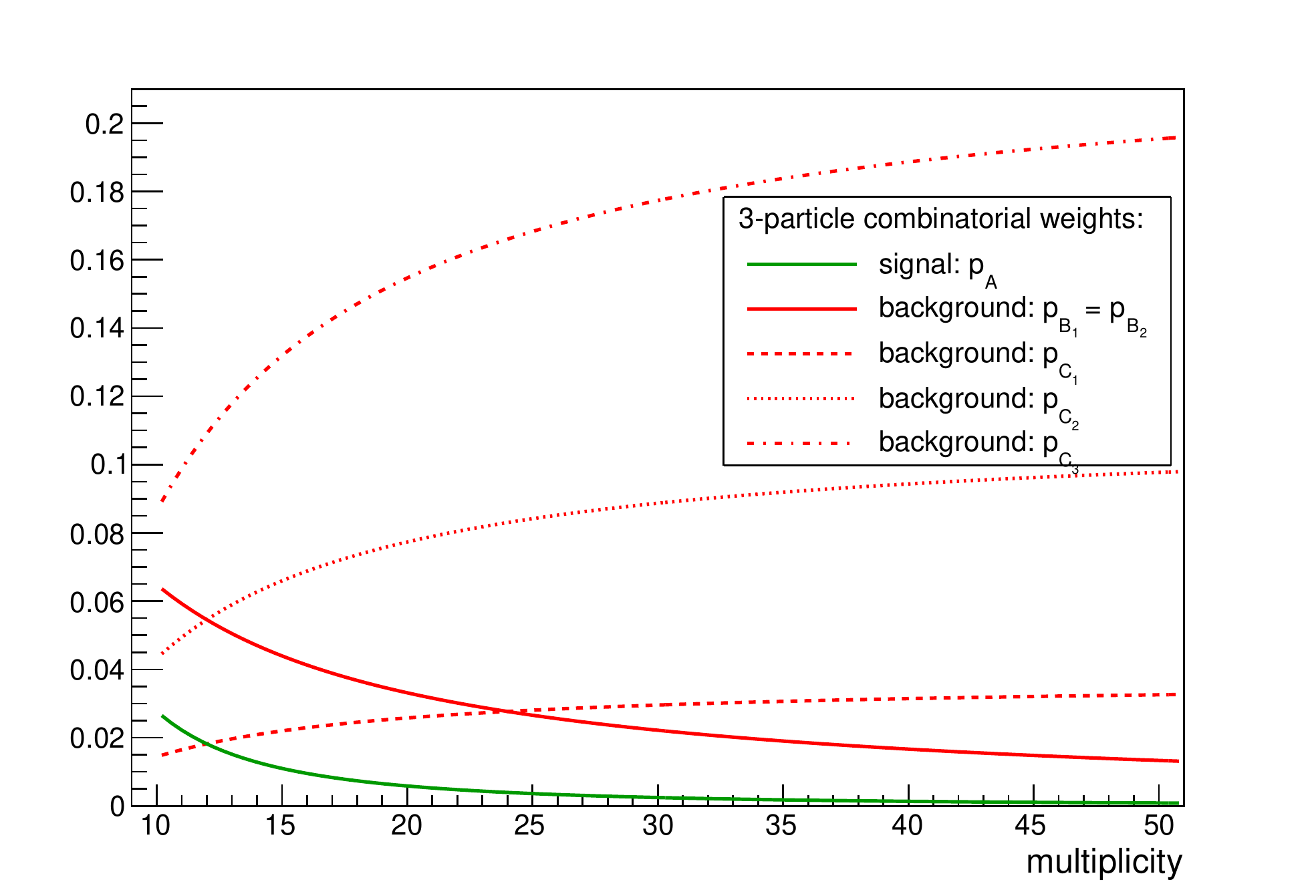}
		\caption{Multiplicity dependence of 6 combinatorial weights for 3-particle correlations: $p_A$, $p_{B_1}$, $p_{B_2}$, $p_{C_1}$, $p_{C_2}$ and $p_{C_3}$ (see the main text for their definitions and further explanation). The contribution from genuine 3-particle correlations (signal) is weighted with $p_A$ (green line) and already for $M\approx 50$ this contribution becomes negligible. The contribution from genuine 2-particle correlations is also suppressed for large multiplicities (solid red line). The remaining weights, $p_{C_1}$, $p_{C_2}$ and $p_{C_3}$, all correspond to independent particle emission, and their relative contribution increases as multiplicity increases. 
		} 
		\label{fig:weights-3p}
	\end{center}
\end{figure}

We can now from the above decomposition read off and write down our final solution for this particular example. If the starting p.d.f. is $f_{xyz}(x,y,z)$, the p.d.f. $w(x,y,z)$ after randomization is given by the following universal result:
\begin{eqnarray}
w(x,y,z) &=& p_A f_{xyz}(x,y,z)\nonumber\\
&+& p_{B_1}\big[f_{xy}(x,y)f_{x}(z) + f_{xy}(x,y)f_{y}(z) + f_{xz}(x,z)f_{x}(y)\nonumber\\
&&{} + f_{xz}(x,z)f_{z}(y) + f_{yz}(y,z)f_{y}(x) + f_{yz}(y,z)f_{z}(x)\big]\nonumber\\
&+& p_{B_2}\big[f_{xy}(x,y)f_{z}(z) + f_{xz}(x,z)f_{y}(y) + f_{yz}(y,z)f_{x}(x)\big]\nonumber\\
&+& p_{C_1}\big[f_{x}(x)f_{x}(y)f_{x}(z) + f_{y}(x)f_{y}(y)f_{y}(z) + f_{z}(x)f_{z}(y)f_{z}(z)\big]\nonumber\\
&+& p_{C_2}\big[f_{x}(x)f_{x}(z)f_{y}(y) + f_{x}(x)f_{x}(y)f_{z}(z) + f_{y}(y)f_{y}(z)f_{x}(x)\nonumber\\
&&{} + f_{y}(y)f_{y}(x)f_{z}(z) + f_{z}(z)f_{z}(y)f_{x}(x) + f_{z}(z)f_{z}(x)f_{y}(y) \big]\nonumber\\
&+& p_{C_3}f_{x}(x)f_{y}(y)f_{z}(z)\,.
\label{eq:universalSolution-3p}
\end{eqnarray}
This result is universal because the 6 combinatorial weights $p_A$, $p_{B_1}$, $p_{B_2}$, $p_{C_1}$, $p_{C_2}$ and $p_{C_3}$ depend only on multiplicity, while all marginal p.d.f.'s are always by definition calculated in the same way from the starting p.d.f. $f_{xyz}(x,y,z)$ (for instance, $f_{xy}(x,y) \equiv \int f_{xyz}(x,y,z)\,dz$, $f_{x}(x) \equiv \int\!\!\int f_{xyz}(x,y,z)\,dydz$, etc.). Therefore, whatever the starting p.d.f. $f_{xyz}(x,y,z)$ is, one needs to calculate all marginal p.d.f.'s, insert them in the above equation, and the result for the p.d.f. $w(x,y,z)$ of randomized sample follows. We stress again that the example analytic result in Eq.~(\ref{eq:universalSolution-3p}) can be further generalized by including the contributions when mis-identified particles corresponding to the same initial particle type are coupled directly to each other in the 3-particle correlators.

Figure~\ref{fig:weights-3p} shows multiplicity dependence of 6 combinatorial weights for 3-particle correlations: $p_A$, $p_{B_1}$, $p_{B_2}$, $p_{C_1}$, $p_{C_2}$ and $p_{C_3}$. As expected, all weights that correspond either to genuine 3-p correlation ($p_A$) or genuine 2-particle correlations ($p_{B_{1}}$ and $p_{B_{2}}$) diminish rather quickly as multiplicity increases. 

In the next section, we use the clear-cut Toy Monte Carlo study, to confirm the validity of our main result in Eq.~(\ref{eq:universalSolution-3p}). 

\subsubsection{Toy Monte Carlo study for 3-particle case}

In this section we set up a Toy Monte Carlo study, to confirm the validity of our main result in Eq.~(\ref{eq:universalSolution-3p}). We organize a study in such a way that it does not lead to simplifications at any step. For instance, we choose random observables $x$, $y$ and $z$ to have different functional forms of single-variate marginal p.d.f.'s, defined over different sample spaces.

The starting 3-variate p.d.f. $f(x,y,z;M)$ is given by the following expression:
\begin{equation}
f(x,y,z;M) \equiv\frac{100+Mx+2My+3Mz}{6(100+7M)}\,,x\in(0,1),y\in(0,2),z\in(0,3)\,. 
\label{eq:startingPDF-toyMC-3p}
\end{equation}
As defined above, the three stochastic variables are $x$, $y$ and $z$, while $M$ is a parameter that corresponds to multiplicity. The normalization constraint is satisfied for any $M$, i.e.:
\begin{eqnarray}
\int_0^1\!\!\!\int_0^2\!\!\!\int_0^3 f(x,y,z;M) \,dxdydz &=& 1\,,\forall M\,.
\end{eqnarray}
This toy model was carefully designed with the following two key aspects in mind: a) It can be solved analytically; b) All individual terms in Eq.~(\ref{eq:universalSolution-3p}) have different non-vanishing contributions (i.e. there are no underlying symmetries due to which some terms would be identically zero). From the p.d.f. in Eq.~(\ref{eq:startingPDF-toyMC-3p}), we sample three particles $M/3$ times, to obtain event with multiplicity $M$. The true value of average 3-particle correlation (signal), $\langle xyz\rangle_S$, can be obtained with the straightforward calculus. We have obtained:
\begin{eqnarray}
\langle xyz\rangle_S &=& \int_0^1\!\!\!\int_0^2\!\!\!\int_0^3 xyz f(x,y,z;M) \,dxdydz\nonumber\\ 
&=& \frac{75+7M}{100+7M}\,.
\label{eq:trueCorrelation-3p} 
\end{eqnarray}
Now we proceed with the non-trivial part. In each event we randomize the sampled variables, and calculate the average 3-particle correlation $\langle xyz\rangle_{S+B}$ in the randomized sample (signal + background). To get the theoretical result for $\langle xyz\rangle_{S+B}$, in the first step we calculate all marginal p.d.f.'s $f_{xy},f_{xz},f_{yz},f_{x},f_{y}$ and $f_{z}$ from the starting 3-variate p.d.f. $f(x,y,z;M)$ in Eq.~(\ref{eq:startingPDF-toyMC-3p}). For instance, for the 2-variate marginal p.d.f. of $x$ and $y$ we have obtained the following analytic result:
\begin{equation}
f_{xy}(x,y;M) = \frac{200+M(9+2x+4y)}{400+28M}\,.
\end{equation}
We remark that by definition if the starting p.d.f. is normalized, any marginal p.d.f.'s calculated from it are automatically normalized, e.g. we have:
\begin{equation}
\int_0^1\!\!\!\int_0^2 f_{xy}(x,y;M)\,dxdy = 1\,,\forall M\,.
\end{equation}
After we have obtained all marginal p.d.f.'s $f_{xy},f_{xz},f_{yz},f_{x},f_{y}$ and $f_{z}$, we insert them in Eq.~(\ref{eq:universalSolution-3p}), together with universal combinatorial weights for 3-particle case $p_A$, $p_{B_1}$, $p_{B_2}$, $p_{C_1}$, $p_{C_2}$, and $p_{C_3}$, to obtain the analytic result for the p.d.f. of randomized sample $w(x,y,z)$. Finally, with respect to this p.d.f. we calculate the average 3-particle correlation, which now describes analytically both the contributions from signal and from combinatorial background. After some straightforward calculus, we have obtained our final result:
\begin{eqnarray}
\langle xyz\rangle_{S+B} &=& \int_0^1\!\!\!\int_0^2\!\!\!\int_0^3 xyz\, w(x,y,z;M) \,dxdydz\nonumber\\ 
&=& \frac{9000000+2M(-4132500+M(316475+M(258423+4M(6487+192M))))}{3(M\!-\!1)(M\!-\!2)(7M\!+\!100)^3}\,.
\label{eq:finalToyMC-3p} 
\end{eqnarray}
\begin{figure}[h!]
	\begin{center}
		\includegraphics[scale=.5]{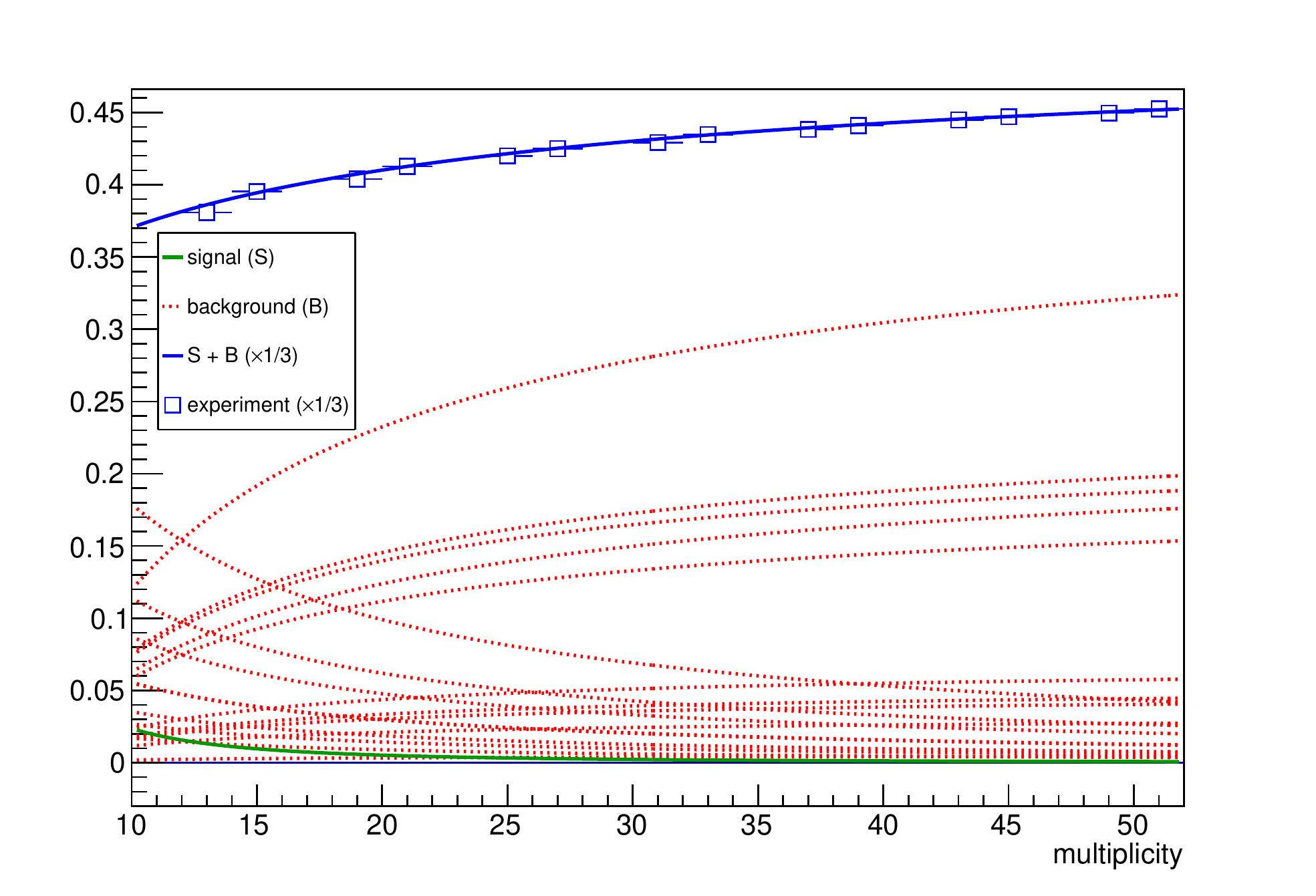}
		\caption{Example analytic description of combinatorial background in 3-particle correlations, when particles are produced in triples, and the final data sample is randomized.} 
		\label{fig:toyMC-3p}
	\end{center}
\end{figure}
These results are shown in Fig.~\ref{fig:toyMC-3p}. With blue markers we have shown the experimental result, i.e. the average $\langle xyz\rangle$ measured in the randomized data sample. Both signal and combinatorial background contribute to this measurement. The analytic expression for the signal contribution is shown with the green curve---this is the result from Eq.~(\ref{eq:trueCorrelation-3p}) weighted with its combinatorial weight $p_{A} = \frac{3!\frac{M}{3}}{M(M-1)(M-2)}$. Red curves represent analytic results for 19 distinct cases of combinatorial background contribution, weighted with corresponding combinatorial weights $p_{B_{1}}, p_{B_{2}}, p_{C_{1}}, p_{C_{2}}$ and $p_{C_{3}}$ (see Eq.~(\ref{eq:universalSolution-3p})). Finally, the blue curve is given by Eq.~(\ref{eq:finalToyMC-3p}), and it is the sum of signal (green curve) and distinct background contributions (19 red curves). This study demonstrates that to great precision the example theoretical result from Eq.~(\ref{eq:universalSolution-3p}) describes the measurement for 3-particle correlation in the randomized data sample, when both signal and combinatorial background contributions are present. 

The result in Eq.~(\ref{eq:universalSolution-3p}) can be further generalized and fine-tuned by adding the terms corresponding to subdominant cases of particle mis-identification, but conceptually the structure of solution is always the same: The relation between original p.d.f. $f$ and its counterpart $w$ in the randomized dataset is universal, in a sense that that relation is determined solely by combinatorial weights $p_i$ which depend only on multiplicity, and marginal p.d.f.'s of $f$.

\section{Summary}
\label{s:Summary}
In summary, we have discussed the role of reflection symmetry, permutation symmetry, frame independence, and relabeling of particle indices in the cumulant expansion. The conclusions are general and applicable to cumulants of any stochastic variables, however, the primary aim was to further clarify the properties of the recently proposed event-by-event cumulants of azimuthal angles. For the first time, the analytic solutions for the contribution of combinatorial background in the measured 2- and 3-particle correlations were derived. The main result is the demonstration that these solutions for the combinatorial background are universal---they can be written generically in terms of multiplicity-dependent combinatorial weights and marginal probability density functions of starting multivariate distribution. Finally, the new general results between expectation values of multiparticle azimuthal correlators, and flow amplitudes and symmetry planes were presented.

\begin{acknowledgements}
This project has received funding from the European Research Council (ERC) under the European Union’s Horizon 2020 research and innovation programme (grant agreement No 759257).	
\end{acknowledgements}

\appendix

\section{Expectation values}
\label{a:Expectation-values}

In this appendix, we summarize the theoretical expectation values of all estimators used in this work. Since our estimators are defined in terms of sums, where each summand has the same generic expression, their expectation values are trivially related to the expectation values of summands, i.e. the fundamental building blocks in the sums. Therefore, we derive first the theoretical expectation values of summands, which we denote as fundamental observables, and from them trivially deduce the expectation values of high-level compound observables. In all derivations we assume that only anisotropic flow correlations are present, therefore for an event with multiplicity $M$ we can use $f(\varphi_1,\varphi_2,\ldots,\varphi_M) = f(\varphi_1)f(\varphi_2)\cdots f(\varphi_M)$, where every single-particle p.d.f. on the right-hand side is parameterized with the same normalized Fourier series (see introductory part of Ref.~\cite{Bilandzic:2010jr} for further details).

\subsection{Fundamental observables}
\label{aa:Fundamental-observables}

For completeness sake and to unify the notation, we first reproduce the theoretical expectation value of the $k$-particle azimuthal correlator evaluated in harmonics $n_1, \ldots, n_k$, in a data sample containing $M$ azimuthal angles ($k\leq M$): 
\begin{eqnarray}
E[e^{i(n_1\varphi_1+\cdots+n_k\varphi_k)}] &\equiv& \left<e^{i(n_1\varphi_1+\cdots+n_k\varphi_k)}\right>\nonumber\\
&=& \int\cdots\int e^{i(n_1\varphi_1+\cdots+n_k\varphi_k)}f(\varphi_1,\cdots,\varphi_M)\,d\varphi_1 \ldots d\varphi_M\nonumber\\
&=& \int\cdots\int e^{in_1\varphi_1}\cdots e^{in_k\varphi_k}f(\varphi_1)\cdots f(\varphi_M)\,d\varphi_1 \ldots d\varphi_M\nonumber\\
&=& \left(\prod_{i=1}^{k}\int e^{in_i\varphi_i}f(\varphi_i)\,d\varphi_i\right)\times
\left(\prod_{i=k+1}^{M}\int f(\varphi_i)\,d\varphi_i\right)\nonumber\\
&=& \left(\prod_{i=1}^{k} v_{n_i}e^{in_i\Psi_{n_i}}\right)\times 1\nonumber\\
&=& v_{n_1}\cdots v_{n_k}e^{i(n_1\Psi_{n_1}+\cdots+n_k\Psi_{n_k})}\,.
\label{eq:generalResult}
\end{eqnarray}
This very important analytic result was first derived in Ref.~\cite{Bhalerao:2011yg}. From the above general result we can read off directly the following specific results:
\begin{eqnarray}
E[\cos n\varphi_i] &=& v_n\cos\Psi_n\,,\quad\forall n,i \label{eq:cos-i}\,,\\  
E[\sin n\varphi_i] &=& v_n\sin\Psi_n\,,\quad\forall n,i \label{eq:sin-i}\,,\\
E[\cos[n(\varphi_i-\varphi_j)]] &=& v_n^2\,,\quad\forall n,i,j\ (i\neq j) \label{eq:cos-ij}\,,\\  
E[\sin[n(\varphi_i-\varphi_j)]] &=& 0 \,,\quad\forall n,i,j\ (i\neq j) \label{eq:sin-ij}\,.
\end{eqnarray}

By following the similar procedure, we have derived the following new general results for additional expectation values of interest:
\begin{eqnarray}
E[\cos^{2p}n\varphi_i] &=& \frac{1}{2^{2p-1}}\left[\frac{1}{2}\binom{2p}{p} + \sum_{k=0}^{p-1}\binom{2p}{k}v_{2(p-k)n}\cos[2(p-k)n\Psi_{2(p-k)n}] \right]\,,\nonumber\\
E[\cos^{2p-1}n\varphi_i] &=& \frac{1}{2^{2p-2}}\sum_{k=0}^{p-1}\binom{2p-1}{k}v_{(2p-2k-1)n}\cos[(2p-2k-1)n\Psi_{(2p-2k-1)n}]\,,\nonumber\\  
E[\sin^{2p}n\varphi_i] &=& \frac{1}{2^{2p-1}}\left[\frac{1}{2}\binom{2p}{p} + \sum_{k=0}^{p-1}(-1)^{p-k}\binom{2p}{k}v_{2(p-k)n}\cos[2(p-k)n\Psi_{2(p-k)n}] \right]\,,\nonumber\\
E[\sin^{2p-1}n\varphi_i] &=& \frac{1}{2^{2p-2}}\sum_{k=0}^{p-1}(-1)^{p+k-1}\binom{2p-1}{k}v_{(2p-2k-1)n}\sin[(2p-2k-1)n\Psi_{(2p-2k-1)n}]\,.
\label{eq:general-eqs-case-2}
\end{eqnarray}
In the derivation, we have used result 1.320 from~Ref.~\cite{zwillinger2007table} in combination with the orthogonality relations of trigonometric functions. As an example, we can now read off easily the following specific cases: 
\begin{eqnarray}
E[\cos^2 n\varphi_i] &=& \frac{1}{2}\left[1 + v_{2n}\cos (2n\Psi_{2n})\right]\,,\quad\forall n,i\label{sin-isin-j}\,,\\  
E[\cos^3 n\varphi_i] &=& \frac{1}{4}\left[3v_{n}\cos (n\Psi_{n}) + v_{3n}\cos (3n\Psi_{3n})\right]\,,\quad\forall n,i\label{sin-isin-j}\,,\\  
E[\sin^2 n\varphi_i] &=& \frac{1}{2}\left[1 - v_{2n}\sin (2n\Psi_{2n})\right]\,,\quad\forall n,i\,,\\
E[\sin^3 n\varphi_i] &=& \frac{1}{4}\left[3v_{n}\cos (n\Psi_{n}) - v_{3n}\sin (3n\Psi_{3n})\right]\,,\quad\forall n,i\label{sin-isin-j}\,.  
\end{eqnarray}
In fact, the results in Eqs.~(\ref{eq:general-eqs-case-2}) suffice to decipher the expectation values of most general products, because:
\begin{equation}
E[(\prod_{i=1}^{k_c}\cos^{p_i}n_i\varphi_i)(\prod_{j=k_c+1}^{k_s}\sin^{r_j}m_j\varphi_j)] = \prod_{i=1}^{k_c}E[\cos^{p_i}n_i\varphi_i] \times \prod_{j=k_c+1}^{k_s}E[\sin^{r_j}m_j\varphi_j]\,.
\end{equation}
For instance:
\begin{eqnarray}
E[\cos n_1\varphi_i\cos n_2\varphi_j] &=& v_{n_1}v_{n_2}\cos\Psi_{n_1}\cos\Psi_{n_2}\,,\quad\forall n_1,n_2,i,j\ (i\neq j)\label{cos-icos-j}\,,\\  
E[\sin n_1\varphi_i\sin n_2\varphi_j] &=&  v_{n_1}v_{n_2}\sin\Psi_{n_1}\sin\Psi_{n_2}\,,\quad\forall n_1,n_2,i,j\ (i\neq j)\label{sin-isin-j}\,, 
\end{eqnarray}
We now demonstrate that the expectation values of all high-level observables can be reduced to these fundamental ones.

\subsection{Compound observables}
\label{aa:Compound-observables}
The expectation value of any compound observable can be now trivially deduced from the fundamental expectation values presented in the previous section. For instance, for single-event average 2-particle correlation, $\langle 2\rangle$, we use the following statistics build from azimuthal angles $\varphi_1, \ldots, \varphi_M$ ($M$ is multiplicity of an event):
\begin{equation}
\langle 2\rangle \equiv \frac{1}{M(M-1)}\sum_{i\neq j}^M \cos(\varphi_i-\varphi_j)\,.\\
\end{equation}
Its expectation value is:
\begin{eqnarray}
E[\langle 2\rangle] &=& \frac{1}{M(M-1)}\sum_{i\neq j}^M E[\cos(\varphi_i-\varphi_j)]\nonumber\\
&=& \frac{1}{M(M-1)}\sum_{i\neq j}^M v^2\nonumber\\
&=& \frac{1}{M(M-1)} M(M-1) v^2\nonumber\\
&=& v^2\,.
\end{eqnarray}
%

\nocite{*}
\bibliographystyle{spphys}  
\bibliography{bibliography} 

\begin{thebibliography}{10}
\providecommand{\url}[1]{{#1}}
\providecommand{\urlprefix}{URL }
\expandafter\ifx\csname urlstyle\endcsname\relax
  \providecommand{\doi}[1]{DOI \discretionary{}{}{}#1}\else
  \providecommand{\doi}{DOI \discretionary{}{}{}\begingroup
  \urlstyle{rm}\Url}\fi

\bibitem{Heinz:2013th}
U.~Heinz, R.~Snellings, Ann. Rev. Nucl. Part. Sci. \textbf{63}, 123 (2013).
\newblock \doi{10.1146/annurev-nucl-102212-170540}

\bibitem{Braun-Munzinger:2015hba}
P.~Braun-Munzinger, V.~Koch, T.~Schäfer, J.~Stachel, Phys. Rept. \textbf{621},
  76 (2016).
\newblock \doi{10.1016/j.physrep.2015.12.003}

\bibitem{Busza:2018rrf}
W.~Busza, K.~Rajagopal, W.~van~der Schee, Ann. Rev. Nucl. Part. Sci.
  \textbf{68}, 339 (2018).
\newblock \doi{10.1146/annurev-nucl-101917-020852}

\bibitem{doi:10.1143/JPSJ.17.1100}
R.~Kubo, Journal of the Physical Society of Japan \textbf{17}(7), 1100 (1962).
\newblock \doi{10.1143/JPSJ.17.1100}.
\newblock \urlprefix\url{https://doi.org/10.1143/JPSJ.17.1100}

\bibitem{Ollitrault:1992bk}
J.Y. Ollitrault, Phys. Rev. \textbf{D46}, 229 (1992).
\newblock \doi{10.1103/PhysRevD.46.229}

\bibitem{Voloshin:1994mz}
S.~Voloshin, Y.~Zhang, Z. Phys. \textbf{C70}, 665 (1996).
\newblock \doi{10.1007/s002880050141}

\bibitem{Heiselberg:1996xa}
H.~Heiselberg, A.P. Vischer, Phys. Rev. C \textbf{55}, 874 (1997).
\newblock \doi{10.1103/PhysRevC.55.874}

\bibitem{Heinz:2004pv}
U.W. Heinz, A.~Sugarbaker, Phys. Rev. C \textbf{70}, 054908 (2004).
\newblock \doi{10.1103/PhysRevC.70.054908}

\bibitem{Borghini:2000sa}
N.~Borghini, P.M. Dinh, J.Y. Ollitrault, Phys. Rev. C \textbf{63}, 054906
  (2001).
\newblock \doi{10.1103/PhysRevC.63.054906}

\bibitem{Borghini:2001vi}
N.~Borghini, P.M. Dinh, J.Y. Ollitrault, Phys. Rev. \textbf{C64}, 054901
  (2001).
\newblock \doi{10.1103/PhysRevC.64.054901}

\bibitem{Bilandzic:2010jr}
A.~Bilandzic, R.~Snellings, S.~Voloshin, Phys. Rev. \textbf{C83}, 044913
  (2011).
\newblock \doi{10.1103/PhysRevC.83.044913}

\bibitem{Bilandzic:2013kga}
A.~Bilandzic, C.H. Christensen, K.~Gulbrandsen, A.~Hansen, Y.~Zhou, Phys. Rev.
  \textbf{C89}(6), 064904 (2014).
\newblock \doi{10.1103/PhysRevC.89.064904}

\bibitem{Mordasini:2019hut}
C.~Mordasini, A.~Bilandzic, D.~Karako\c{c}, S.F. Taghavi, Phys. Rev. C
  \textbf{102}(2), 024907 (2020).
\newblock \doi{10.1103/PhysRevC.102.024907}

\bibitem{ALICE:2021klf}
S.~Acharya, et~al., Phys. Rev. Lett. \textbf{127}(9), 092302 (2021).
\newblock \doi{10.1103/PhysRevLett.127.092302}

\bibitem{Bilandzic:2021rgb}
A.~Bilandzic, M.~Lesch, C.~Mordasini, S.F. Taghavi, Phys. Rev. C
  \textbf{105}(2), 024912 (2022).
\newblock \doi{10.1103/PhysRevC.105.024912}

\bibitem{Cowan:1998ji}
G.~Cowan, \emph{{Statistical data analysis}} (1998)

\bibitem{Bhalerao:2011yg}
R.S. Bhalerao, M.~Luzum, J.Y. Ollitrault, Phys. Rev. C \textbf{84}, 034910
  (2011).
\newblock \doi{10.1103/PhysRevC.84.034910}

\bibitem{zwillinger2007table}
D.~Zwillinger, A.~Jeffrey, \emph{Table of Integrals, Series, and Products}
  (Elsevier Science, 2007).
\newblock \urlprefix\url{https://books.google.de/books?id=aBgFYxKHUjsC}

\end{thebibliography}

\end{document}